# Windowed Carbon Nanotubes for Efficient $CO_2$ Removal from Natural Gas

*Hongjun Liu,[†] Valentino R. Cooper,[‡] Sheng Dai,[†,§] and De-en Jiang*[,†]*

[†]Chemical Sciences Division and [‡]Materials Science and Technology Division, Oak Ridge National Laboratory, Oak Ridge, Tennessee 37831, and [§]Department of Chemistry, University of Tennessee, Knoxville, Tennessee 37966

*To whom correspondence should be addressed. E-mail: jiangd@ornl.gov.

**ABSTRACT**

We demonstrate from molecular dynamics simulations that windowed carbon nanotubes can efficiently separate $CO_2$ from the $CO_2/CH_4$ mixture, resembling polymeric hollow fibers for gas separation. Three $CO_2/CH_4$ mixtures with 30%, 50% and 80% $CO_2$ are investigated as a function of applied pressure from 80 to 180 bar. In all simulated conditions, only $CO_2$ permeation is observed; $CH_4$ is completely rejected by the nitrogen-functionalized windows or pores on the nanotube wall in the accessible timescale, while maintaining a fast diffusion rate along the tube. The estimated time-dependent $CO_2$ permeance ranges from $10^7$ to $10^5$ GPU (gas permeation unit), compared with ~100 GPU for typical polymeric membranes. $CO_2/CH_4$ selectivity is estimated to be ~$10^8$ from the difference in free-energy barriers of permeation. This work suggests that a windowed carbon nanotube can be used as a highly efficient medium, configurable in hollow-fiber-like modules, for removing $CO_2$ from natural gas.

**TOC graphic**

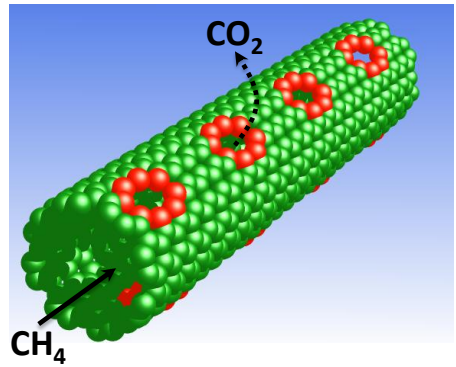

**KEYWORDS:** Natural gas sweetening, carbon dioxide removal, windowed carbon nanotube, molecular dynamics, nanofluidics, nanoporous material.

Natural gas has become increasingly important as a source of cleaner energy than coal and oil as unconventional sources such as shale gas are being explored.[1-2] Natural gas usually contains excess $CO_2$ that must be removed to meet pipeline and heating value specifications. $CO_2/CH_4$ separations are also vital in landfill gas recovering and enhanced oil recovery. The most widely used process to sweeten natural gas employs the alkanolamine aqueous solution to selectively absorb $CO_2$ and $H_2S$ from sour natural gas streams.[3] The process is energy intensive, as the amine solution needs to be heated up to recover the acid gases; in addition, amine is prone to degradation and evaporation loss.

Due to their intrinsic energy efficiency, membranes have been increasingly adopted in industry for gas separations.[4-5] Polymeric membranes in a hollow-fiber module have become the de facto standard for industrial gas separation by membranes, due to the ease in solution processing of a polymer into spiral wound hollow fibers.[6-8] A polymer membrane's performance is usually bound by a tradeoff between permeability and selectivity.[9] Although cellulosic acetate has been a workhorse in industrial membrane separations, novel polymers continue to attract researchers' interest, in a goal to break the upper bound; thermally rearranged polymers[10-11] and PIMs (polymers of intrinsic microporosity)[12-13] are such examples.

The permeance or productivity of a membrane is inversely proportional to the thickness of the membrane. This relationship led us to propose porous graphene for gas separations as a one-atom-thick membrane.[14] After this proof of concept, many more computational studies of interesting separations were followed.[15-25] For example, it has been shown that helium isotope separation can be achieved through the nanoporous graphene.[16, 20-21] Beyond gas separations, Cohen-Tanugi and Grossman[22] showed the great potential of nanoporous graphene for water desalination, well above the conventional reverse osmosis membrane. More excitingly, Bunch and coworkers have experimentally demonstrated the gas-separating power of a porous graphene by purposefully creating subnanometer pores on a graphene sheet and observing different leaking rates for various gases;[26] this elegant experiment convincingly showed selective passing-through of $H_2$ molecules.

To move porous graphene towards practical gas separation, a configuration such as the polymeric hollow fibers is highly desirable. Moreover, porous graphene's ability for the important $CO_2/CH_4$ separation, however, has not been demonstrated. A close analog to a hollow fiber would be a carbon nanotube. Fast transport of gas molecules through carbon nanotubes has been demonstrated before.[27-30] In addition, aligned carbon nanotubes embedded in a polymer matrix have been shown to yield high-flux transport of liquids.[29, 31]

To combine the advantages of carbon nanotubes, porous graphene, and polymeric hollow fibers, we propose a novel system − windowed carbon nanotube – for gas separation. It comprises a carbon nanotube with molecular size windows or pores on the wall. Such a system has three advantages: (1) the pore size and functionality can be tuned for selective gas separation, thereby yielding high selectivity; (2) the atom-thin wall will provide high permeance, while the tube configuration can be aligned and packed together into a module, just like polymeric hollow fibers used in gas separation; (3) the tube can provide a fast-transport channel for the retentate gas. In this letter, we demonstrate the capacity of windowed nanotubes with nitrogen doped pores for high-permeance and high-selectivity of $CO_2$ removal from natural gas using molecular dynamics simulations.

Our windowed nanotube model includes two coaxial single-wall carbon nanotubes. The outer one is pristine and the inner one is windowed. The outer tube is used to conveniently define a clear boundary and not absolutely necessary; the inner tube with its windows provide the gas-separating capability. This coaxial setup can allow us to get an estimate of the pressure drop as the gas molecules diffuse across the wall of the inner tube into the space between the two tubes. The windows or nanopores of the inner tube are prepared as follows: two neighboring benzene rings are removed, and four dangling bonds are saturated by hydrogen, while the other four dangling bonds together with their carbon atoms are replaced with nitrogen atoms (Figure 1a). Hereafter we refer to this pore as the 4N4H pore, on which all the simulations in this paper are based. Porous graphene with this pore was previously shown to be effective for $H_2/CH_4$ separation.[14] Pairs of the 4N4H pores opposite each other are created along the tube, with a density being at about one pair of windows per 5 nm of the tube length. All-atom models are used for gas molecules. The force field parameters are summarized and tabulated in the Supporting Information (SI).

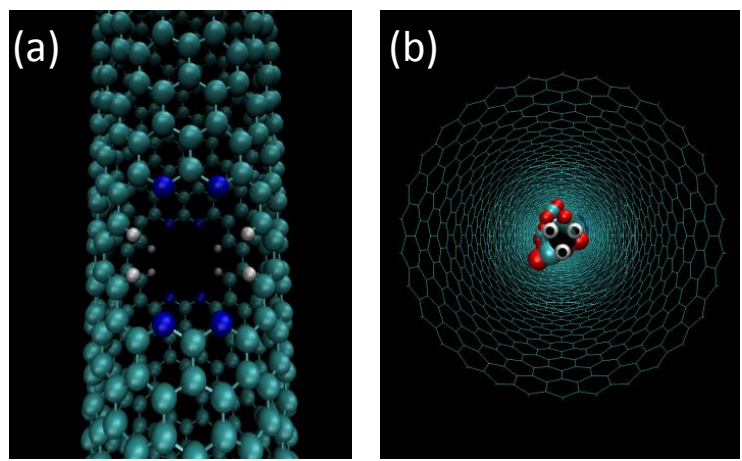

Figure 1. (a) The 4N4H windows or pores on the wall of the inner tube. (b) The initial setup of the simulation where $CO_2/CH_4$ gas mixture is inside the windowed inner tube; on the outside is a pristine tube. The following color code is adopted throughout: carbon (cyan); oxygen (red); hydrogen (white); nitrogen (blue).

The molecular dynamics simulations are performed in the NVT ensemble at 300 K and on a periodic box with the LAMMPS package.[32] Initially, an appropriate amount of $CH_4$ and $CO_2$ mixture based on the molar fraction is loaded into the inner tube (see Figure 1b) for a total length of 20 nm. The number of permeate molecules passing through the nanopores into the space between the two tubes is monitored to quantify flux and permeance; metadynamics simulations[33] are performed to compute free energy of $CH_4$ and $CO_2$ passing through the windows, using the PLUMED plugin[34] with the LAMMPS package. See SI for details of the simulation.

We start with the (14,0) carbon nanotube for the inner tube, (30,0) for the outer, and an initial loading of four total molecules per nanometer length of the tube. This initial configuration is denoted as 14-30-4. Figure 2 shows $CO_2$ permeation in the 14-30-4 configuration with a mixture of 80% $CO_2$ and 20% $CH_4$ inside the inner tube initially. One can see from Figure 2a that $CO_2$ can readily permeate through the nanopores; in contrast, $CH_4$ never passes through during the 16-ns simulation time, indicating a molecular-sieve-like separation. The representative snapshot near the end of the 16-ns simulation (Figure 2b) vividly demonstrates the separation efficiency of the inner windowed nanotube.

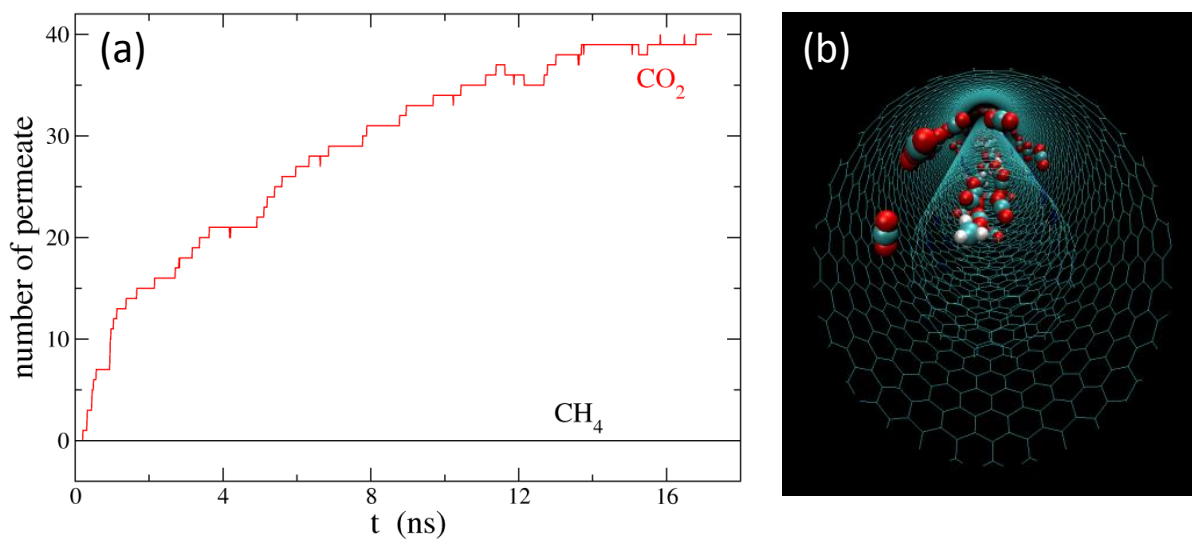

Figure 2. (a) Number of permeate molecules with time for the 14-30-4 initial configuration where the inner windowed tube is (14,0), the outer pristine one is (30,0), and the starting loading of 4 molecules per nm inside the inner tube. The gas mixture is composed of 80% $CO_2$ and 20% $CH_4$; the initial total pressure inside the inner tube is about 175 bar. (b) A representative snapshot around t=16 ns.

To elucidate the mechanism of $CO_2$ permeation, we have characterized one single $CO_2$ molecule passing event in Figure 3a. The trajectory reveals a fast dynamics of $CO_2$ permeation. $CO_2$ passes through the nanopore without significantly binding with it, implying a low permeation energy barrier. It is also evident that there are dozens of failed passing attempts before a successful permeation event. When $CO_2$-passing does happen (Figure 3b), the $CO_2$ molecule spends some time orientating itself from the preferred parallel configuration so that it can perpendicularly pass the pore. Once entering the outer channel, $CO_2$ spends almost all of its time in between the two tubes, where the energy minimum is located.

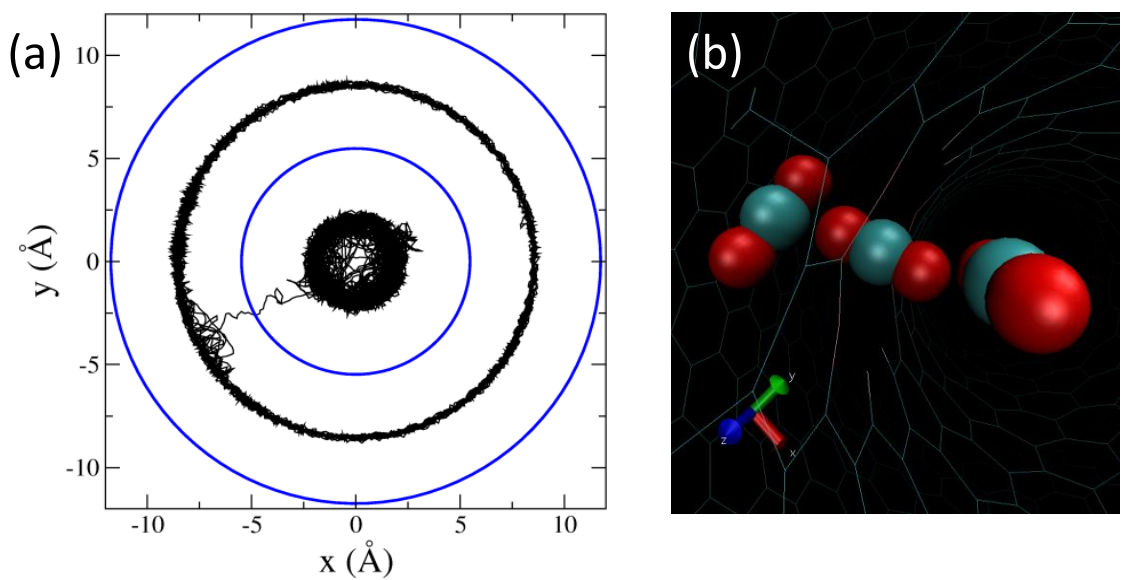

Figure 3. (a) Projected trajectory of a single event of $CO_2$ passing-through in the 14-30-3 configuration where the inner windowed tube is (14,0), the outer pristine one is (30,0), and the starting loading of 3 molecules per nm inside the inner tube. The gas mixture is composed of 80% $CO_2$ and 20% $CH_4$; the initial total pressure inside the inner tube is about 130 bar. The 1-ns time sequence is from the inner tube to the outer channel. Blue, carbon nanotubes; black, $CO_2$ trajectory. (b) A close-up of a single $CO_2$ molecule passing through the 4N4H pore from right to left.

The N4H4 pore on the wall of the inner tube (Figure 1a) has an estimated size of 3.0 Å.[14] $CH_4$ has a kinetic diameter of 3.8 Å, so it is expected that $CH_4$ cannot pass through the pore and therefore remain inside the inner tube. However, $CO_2$'s kinetic diameter is about 3.3 Å, slightly larger than the pore size. Our simulations here show that $CO_2$ can easily pass through the 4N4H pores at 300 K, suggesting that factors other than size play a role for the high $CO_2/CH_4$ separation efficiency. From a recent simulation of nitrogen-doped porous carbon for $CO_2$ adsorption, it has been shown that pyridinic nitrogen, such as those in the N4H4 pore, has a partial negative charge that attracts $CO_2$.[35] This attraction would in fact lower the barrier of passing through. To confirm this hypothesis, we performed metadynamics simulations to obtain the free energy profiles of passing through (Figure 4), and the free energy barriers are 5.5 kcal/mol for $CO_2$ and 17 kcal/mol for $CH_4$. The transition state for $CH_4$ is at the center of the pore, but for $CO_2$ it is in fact 1.5 Å from the center of the pore. This is because the barrier for $CO_2$ originates from the energy penalty to reorient $CO_2$ from the parallel configuration in order to pass through the pore perpendicularly (Figure 3b).

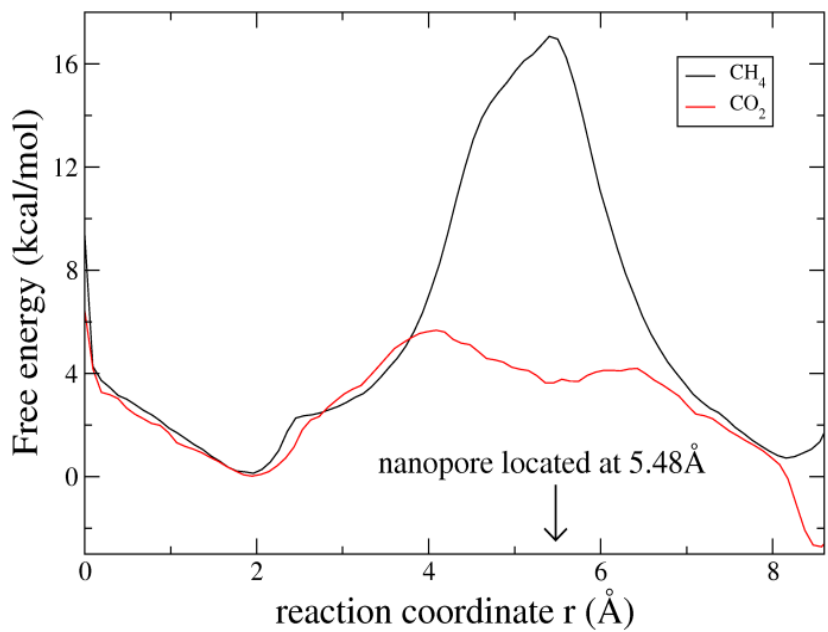

Figure 4. Free energy profiles of passing-through from metadynamics simulation with the radial distance from the axis of nanotubes as the reaction coordinate (or collective variable); the system is in the 14-30-2 configuration (total initial pressure inside the inner tube is about ~88 bar) and the gas mixture is composed of 50% $CO_2$ and 50% $CH_4$.

Based on the free energy barriers from Figure 4, we can now use the Arrhenius equation to estimate the $CO_2$ permeation rate and the $CO_2/CH_4$ selectivity. Assuming a prefactor of $10^{13}$ s$^{-1}$, we obtain a permeation rate of ~1 ns$^{-1}$ for $CO_2$, in good agreement with the rate we see in our MD simulations (for example, in Figure 2a). In addition, the difference in free energy barriers yields a selectivity on the order of $10^8$ for $CO_2/CH_4$. To assess the pressure effect on the selectivity, we also compute the free energy barriers at 300 K at low (~4 bar) and high (~175 bar) initial pressures inside the inner tube. In both cases (see Figures S1 in the SI), a similar selectivity of $10^8$ is obtained for $CO_2/CH_4$, suggesting that the high selectivity can be maintained from the ambient up to 175 bar. For comparison, polymer membranes usually have a $CO_2/CH_4$ selectivity of 10 to 60.[5]

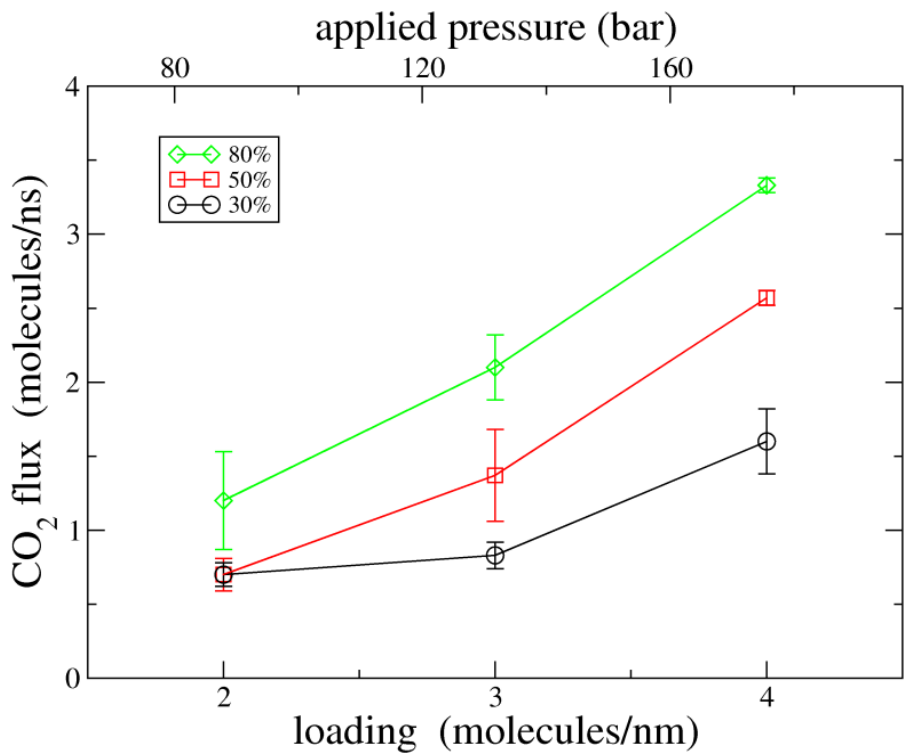

Figure 5. Loading/pressure effect on the flux of $CO_2$ for different $CO_2$ concentrations in the $CO_2/CH_4$ mixture. The inner windowed tube is (14,0), and the outer one is (30,0). No $CH_4$ permeation was seen during the simulation timeframe (10 ns).

To better quantify the separation performance by the windowed nanotubes, we calculated the average $CO_2$ flux over a 10-ns timeframe as a function of loading and $CO_2$ molar fraction in the mixture (see Figure 5). For all the initial configurations considered in Figure 5, we found only $CO_2$ permeation through the inner tube while not a single $CH_4$ passing event was observed, indicating the robustness of the 4N4H pore for highly selective $CO_2/CH_4$ separation. Figure 5 also shows that $CO_2$ flux increases with the loading or applied pressure for the same concentration in the mixture, as higher pressure leads to greater driving force to permeate.

We next estimate the $CO_2$ permeance through the windowed inner tube. Permeance is the flux normalized by the pressure drop across a membrane. The time-dependent flux is obtained by taking the derivative of the number profile of permeate $CO_2$ with time (such as in Figure 2a) fitted to an analytical expression (see SI for details). With the coaxial double-tube setup (Figure 1b), we can straightforwardly estimate time-dependent pressure difference inside and outside the inner tube by the ideal-gas law (see Figure S2 in SI). The time-dependent $CO_2$ permeance is presented in Figure 6 for three different $CO_2$ concentrations with the same total pressure. As expected, higher molar fraction of $CO_2$ leads to higher $CO_2$ permeance. The decrease in permeance with time is due to the decrease in the driving force as more $CO_2$ moves from inside the inner tube to between the tubes. The time-dependent $CO_2$ permeance ranges from $10^7$ to $10^5$ GPU.[36] This permeance is several orders of magnitude greater than that of the typical polymer membranes which usually have a $CO_2$ permeance of ~100 GPU for the $CO_2/CH_4$ mixture.[11] Although over 1000 GPU permeance of $CO_2$ can be achieved now for some advanced polymer membranes,[5] the $CO_2/CH_4$ selectivity they have is usually < 50. In addition, the $CO_2$ permeance of the windowed carbon nanotube can in principle be higher, as we simulated a scenario of low pore density at 0.12 nm$^{-2}$ for the (14,0) inner tube; this density and hence the permeance still have room for improvement.

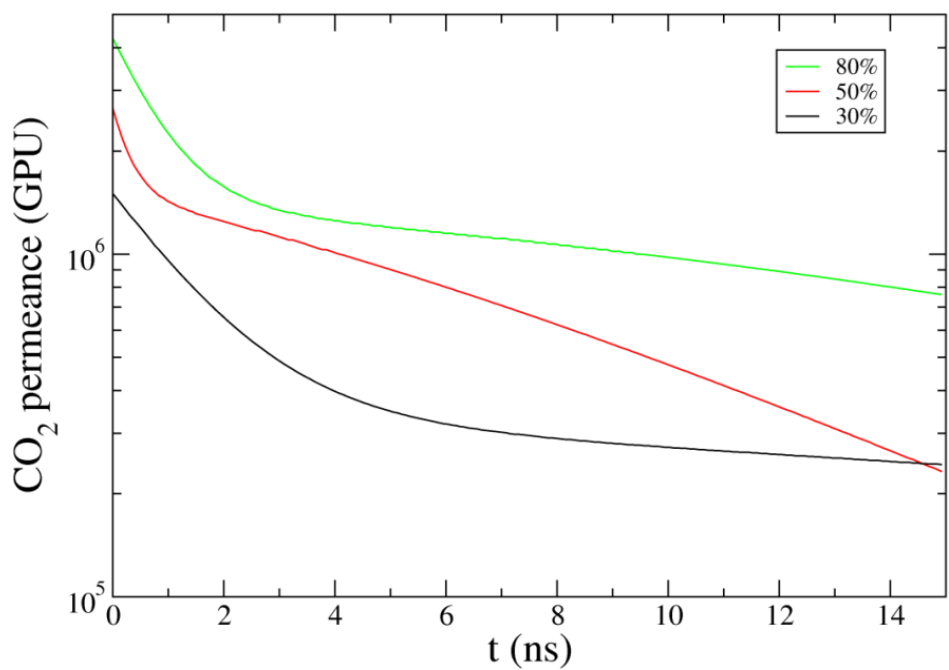

Figure 6. Time-dependent $CO_2$ permeance in the 14-30-4 configuration for different molar fractions of $CO_2$ in the $CO_2/CH_4$ mixture.

To examine the effect of the size of the outer pristine tube on separation of $CO_2$, we varied the outer-tube size so that the spacing between the two tubes changes from 5.48 to 7.83 Å. Figure 7 shows the number of the $CO_2$ permeates with time. One can see that the rate of $CO_2$ permeation shows a dramatic increase when the spacing between the tubes increases from 5.48 Å to 6.26 Å. The rate does not change significantly between 6.26 and 7.83 Å, indicating that high flux of $CO_2$ can be achieved as long as the outer tube size exceeds some threshold. This threshold mainly reflects the need of having an enough space to accommodate $CO_2$.

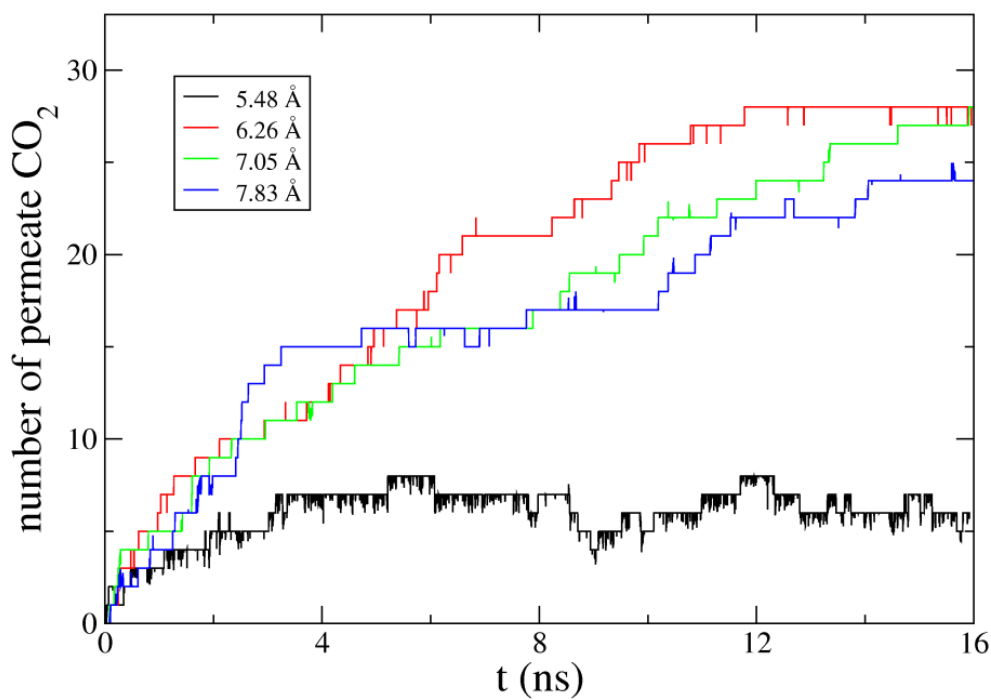

Figure 7. The effect of the outer tube size on $CO_2$ permeation. The caption shows the spacing between the outer tube and the inner windowed (14,0) tube. $CO_2$ molar fraction is at 50% in the $CO_2/CH_4$ mixture; initial gas loading is at 4 per nm.

Besides highly selective and high-flux transport of the permeate through the barrier layer, another important factor in membrane separation is the fast transport of the retentate. After $CO_2$ removal, the remaining $CH_4$ needs to be transported out. To test if fast transport of $CH_4$ inside the windowed inner tube can be achieved, we compared $CH_4$ diffusion in the windowed (14,0) nanotube and in the pristine nanotube at different loadings (see Figure S3 in SI). Although $CH_4$ diffusion becomes slower in the windowed nanotube, it is still one order of magnitude greater than that in zeolites.[30] The retarded diffusion in the windowed nanotube is probably due to the wall roughness caused by the pores.

It would be challenging to make windowed nanotubes with narrowly distributed nanopores. However, recent experimental progress strongly suggests its feasibility. Koenig, *et al.* successfully used UV/ozone etching to obtain the nanoporous graphene as selective molecular sieve for gas separation.[26] Therefore a straightforward idea to create windowed carbon nanotubes would be to use the UV/ozone to etch a carbon nanotube. One can then align the windowed carbon nanotubes for gas-separation tests. Another approach would be to make the porous graphene first either by etching or bottom-up synthesis,[37] and then transform them into nanotubes, as recently demonstrated for few-layer graphene.[38] Although it probably is still far away for windowed nanotubes to be practical for gas separations like polymeric hollow fibers, the high efficiency of $CO_2$/$CH_4$ separation as demonstrated here sure is tantalizing for experimental realization.

To conclude, we have demonstrated from molecular dynamics simulations that windowed carbon nanotubes are able to separate $CO_2$ from the $CO_2$/$CH_4$ mixture with a $CO_2$ permeance several orders of magnitude higher than the typical polymer membrane. The nitrogen-functionalized window or pore rejects $CH_4$ 100% in the simulated conditions and timeframe, indicating a molecular-sieve-like, highly selective separation of $CO_2$ from $CH_4$. Metadynamics simulation shows that the free-energy barrier of $CO_2$ passing-through is small (~5.5 kcal/mol) while that of $CH_4$ is ~ 12 kcal/mol higher, corresponding to a $CO_2$/$CH_4$ selectivity of $10^8$ at room temperature. MD simulations also show that, though hindered by the windows on the wall, the $CH_4$ retentate can still maintain a high diffusion rate inside the tube. Hence this work suggests that windowed carbon nanotubes could be a premium membrane platform for highly selective, high-flux gas separation, in the spirit of the highly successful hollow-fiber configuration of the polymer membrane.

**Acknowledgement.** This work was supported by the Division of Chemical Sciences, Geosciences, and Biosciences, Office of Basic Energy Sciences, U.S. Department of Energy. We thank Dr. Andrew Stack of ORNL for help with metadynamics simulations. This research used resources of the National Energy Research Scientific Computing Center (NERSC), which is supported by the Office of Science of the U.S. Department of Energy under Contract No. DE-AC02-05CH11231.

**Supporting Information Available:** The force field parameters and computational details. This material is available free of charge via the Internet at http://pubs.acs.org.